\documentclass{ijuc}
\usepackage{graphics}
\usepackage{subfigure}
\usepackage{amsmath}

\begin{document}

\title{Reconfigurable Multiple-Valued Logic Function and Sequential Circuit Realizations via Threshold Logic Gates}

\author{Ahmet Unutulmaz\inst{1}\email{ahmet.unutulmaz@marmara.edu.tr}
\and Cem \"{U}nsalan\inst{2} 
}

\institute{Faculty of Engineering, Department of Electrical and Electronics Engineering Marmara University, Istanbul, Turkey
\and
Department of Electrical and Electronics Engineering Yeditepe University, Istanbul, Turkey
}

\maketitle

\begin{abstract}
In this paper, we present a general reconfigurable multiple-valued logic circuit. The proposed architecture is based on threshold logic gate and is compatible with binary logic, which allows a designer to easily integrate multiple valued logic with binary logic. 
We also present a methodology to design sequential circuits.
\end{abstract}


\section{Introduction}

Reconfigurable devices have been extensively used while realizing binary logic circuits. The first reconfigurable logic devices were based on the programmable array logic (PAL) architecture containing an array of AND and OR gates. PAL architectures are capable of executing any Boolean logic expression as a two-level sum-of-products function. However, they require quadratically more number of programmable switches as their logic capacity increases. This limits their scalability. Complex programmable logic devices (CPLDs) aimed to address the scalability issue by integrating multiple PALs on the same die, interconnected through a crossbar. Scalability of reconfigurable devices significantly improved after the introduction of look up table (LUT) based FPGA by Xilinx in 1984. LUT based architectures have much higher area efficiency compared to PALs and CPLDs. Later, and-inverter cone approach was proposed in \cite{parandeh2012rethinking} and field programable transistor array was presented in \cite{tian2017field}. LUT based architectures are the dominant architectures in today's commercial FPGAs. The interested reader should check \cite{boutros2021fpga} for a more detailed review on the evolution of binary FPGA architectures. 

A recent survey on the existing designs and techniques for FPGA logic cells is presented in \cite{rai2021survey}. The paper also examines how emerging technologies and innovative micro-architectures strive to improve performance. The design in \cite{jabeur2011fine} leverages the ambipolar property of double-gate CNTFETs to reduce the delay of reconfigurable cells. An ambipolar device may be configured as either n-type or p-type. An FPGA implementation based on dynamic logic cell architecture is presented in \cite{gaillardon2014novel}, where the logic cells utilize silicon nano-wire ambipolar devices. A spintronic based FPGA implementation alternative to LUT-based CMOS reconfigurable logic is presented in \cite{williams2018architecture}. In \cite{ho2017configurable}, Configurable memristive logic block (CMLB) are build via connecting memristive logic cells using memristive switch matrix cells. The CMLB is then used to construct a memristor-based FPGA architecture.

Traditional binary logic confines each wire to just two distinct voltage levels (representing 0 and 1) in a reconfigurable device. This approach has limitations in representing the inherent real world complexities and uncertainties while enabling efficient computation. Multi-valued logic offers an alternative by expanding beyond binary encoding. Multi-valued logic circuits leverage N distinct voltage levels to represent N different values. This effectively compresses information into a smaller spatial footprint, consequently demanding fewer routing resources and ultimately resulting in a reduction in chip area. This reduction translates directly to cost savings for the circuit designer. In this study, we will interchangeably call multi-valued logic as \textit{Nary} logic as an extension of binary logic. 

Several technologies offer devices that could potentially be used for multi-valued logic applications. A review of such devices, focusing on their operating principles, technologies, and applications, is presented in \cite{andreev2022looking}. Although having high potential in implementation, only a few studies examine how multi-valued logic can be applied in a reconfigurable device. A quaternary look up table (LUT) architecture entirely implemented in CMOS technology is presented in \cite{da2009cmos}. This design necessitated the use of multiple supply voltages and transistors with distinct threshold voltages. The study in \cite{lazzari2010new} introduces a quaternary configurable logic block (CLB) which incorporates dedicated quaternary carry propagation logic and quaternary flip-flops. In \cite{brito2014quaternary} an alternative quaternary LUT architecture, compatible with standard CMOS technology, is introduced. This design employs a single-level multiplexer to achieve reduced output resistance. The study in \cite{chaudhuri2018beyond} implemented a quaternary FPGA architecture on an FDSOI technology. 

Prior works, such as \cite{wagle2021heterogeneous}, employed threshold logic gates (TLG) within binary FPGA architectures. To our knowledge, no studies to date investigated the utilization of TLG for reconfigurable multi-valued logic circuits. Therefore, we propose a framework for implementing multi-valued reconfigurable logic functions in this study. This framework allows the utilization of Nary logic for both input and output signals, while maintaining compatibility with binary logic operations. This approach targets a reduction in wire congestion, potentially leading to lower chip cost. The proposed framework is build upon TLG based Nary decoder and multiplexer circuits. These building blocks are then used to implement multi-valued logic functions and sequential circuits in a reconfigurable manner. 

Layout of the study is as follows. Abstract form of the TLG is introduced in Section~\ref{section:TLG}. Nary decoder and multiplexer circuits based on TLG are introduced in Sections \ref{section:Nary_decoder} and \ref{section:Nary_MUX}, respectively. These building blocks are then used to implement multi-valued logic functions in Section~\ref{section:Nary_Func}. Section~\ref{sec:reconfigurabile_nary} generalizes the idea to reconfigurable circuits. Section~\ref{sec:sequential_nary} focuses on reconfigurable sequential circuit implementations. Finally, Section~\ref{sec:conclussion} provides concluding remarks and future directions.

\section{Threshold Logic Gate}\label{section:TLG}

\begin{figure}[htbp]
    \begin{center}
    	\scalebox{1.3}{\includegraphics{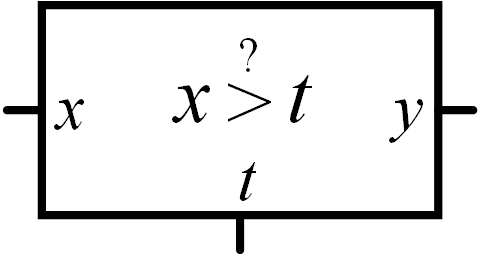}}
  \end{center}
    \caption{Symbol of the TLG.}\label{fig:tlg_sym}
\end{figure}

The framework for implementing Nary reconfigurable logic functions in this study depends on TLG. Therefore, we first introduce it in this section. The TLG used in this study has two inputs as $x$ and $t$. It compares its inputs and outputs a binary logic level 1 if $x>t$, binary logic level 0 otherwise. We will use the symbol in Fig.~\ref{fig:tlg_sym} for the TLG circuit. 

For the implementation details of this Transistor-Level Gate (TLG), we refer the reader to our previous work \cite{unutulmaz2023implementation}. This specific implementation offers a built-in latch, simplifying the design of sequential circuits. However, the framework presented in this paper can also be implemented with other TLG architectures. Throughout the paper, we will employ an abstract representation of the circuits. Therefore, we will focus on the presented framework rather than the underlying transistor-level details.

\section{Nary Decoder Formation via TLG}\label{section:Nary_decoder}

We will benefit from Nary decoders as our first option to implement multiple-valued logic functions. Therefore, we will start with defining the Nary decoder in this section. This decoder works such that only one of its outputs is set to binary logic level 1 and rest of outputs are kept at logic level 0 depending on the Nary input. We will form Nary decoders via TLG by first proposing a one-to-$N$ decoder. Then, we will extend the approach to $M$-to-$N^{M}$ decoder formation, where $M$ is the number of Nary inputs to the decoder. 

\subsection{One-to-$N$ Decoder Formation} \label{sec:one_to_n_decoder}

A one-to-$N$ decoder has one Nary input with $N$ possible values. Output of the decoder is $N$ pins, each being binary logic level 0 or 1 based on the given input. The Nary decoder will have one-hot output as in binary decoders. Hence, only one output pin will be at logic level 1 for the given Nary input.


We can represent the proposed one-to-$N$ decoder mathematically as follows. Let $x$ be the Nary input to the decoder. $N$ outputs of the decoder will be $b_n \in \{0,1\}$ for $n=0,1,\cdots,N-1$. For the given Nary input $x$, only $b_x$ output will be logic level 1. The remaining $b_n$ values will be logic level 0. Hence, one-hot output representation is satisfied.

\begin{figure}[htbp]
    \begin{center}
    	\scalebox{0.8}{\includegraphics{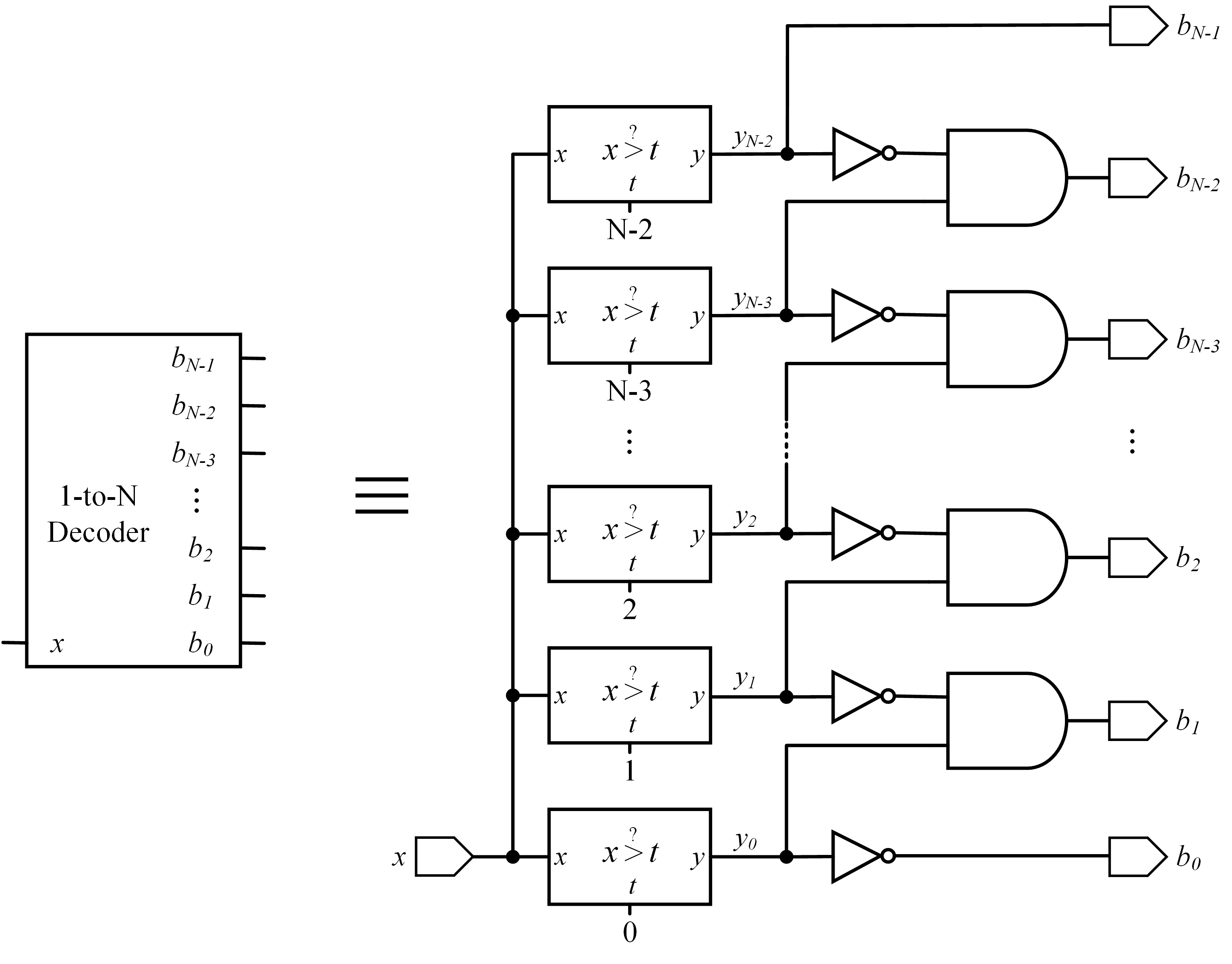}}\label{fig:nary_decoder}
  \end{center}
    \caption{One-to-N decoder. }\label{fig:one_to_N}
\end{figure}

We can realize the one-to-$N$ decoder via TLGs and binary logic gates. We provide the proposed structure in Fig.~\ref{fig:one_to_N}. As can be seen in this figure, operation of the circuit resembles a flash type analog-to-digital converter such that Nary input $x$ is compared against $N-1$ threshold values.

As can be seen in Fig.~\ref{fig:one_to_N}, output of TLGs for the input $x$ will be $y_n \in \{0,1\}$ for $n=0,1,\cdots,N-2$. Binary AND gates in the figure convert outputs $y_n$ to one-hot output representation as $b_n=\bar{y}_n \cdot y_{n-1}$ for $n=0,1,\cdots,N-1$. Here, $\cdot$ represents the binary AND gate. $\bar{y}_n$ indicates the binary NOT of $y_n$. Outputs $y_{-1}$ and $y_{N-1}$ do not exist in the implementation given in Fig.~\ref{fig:one_to_N}. They are included in the representation for mathematical completeness. Therefore, they should be taken as 1 and 0 in implementation, respectively.

Let's consider the ternary decoder with $N=3$ to clarify our discussion. Hence, the circuit in Fig.~\ref{fig:one_to_N} simplifies such that there are two TLGs with threshold values as 1 and 0, respectively. We provide the truth table formed for the ternary decoder in Table~\ref{tab:tt_ternary_odd}. Here, $x$ is the ternary input; $b_2$, $b_1$, and $b_0$ are decoder outputs. As can be seen in the table, only one output has logic level 1 for a given ternary input. 

\begin{table}[htbp]
	\begin{center}
	\begin{tabular}{c|ccc}
		Input &  \multicolumn{3}{c}{Output}\\
		\hline
        $x$	& $b_2$ & $b_1$ & $b_0$\\
		\hline
		    0     & 0         & 0          & 1 \\
		    1     & 0         & 1          & 0 \\
		    2     & 1         & 0          & 0 \\
		\hline
	\end{tabular}
\end{center}
\caption{Truth table for the ternary decoder.}\label{tab:tt_ternary_odd}
\end{table}

\subsection{$M$-to-$N^M$ Decoder Formation} \label{sec:m_to_n_m_decoder}

We can form the $M$-to-$N^M$ decoder following the procedure in the previous section. To do so, we should first decompose it to $M$ one-to-$N$ decoders. Then, we can merge outputs of one-to-$N$ decoders via binary AND gates to reach the final $M$-to-$N^{M}$ decoder. Let's explain the procedure on a simple example. 

Assume that we would like to form a two-to-$N^2$ decoder with two Nary inputs $x_1$, $x_0$ $\in \{0,1,\cdots,N-1\}$. This decoder can be constructed by using two one-to-$N$ decoders. The first decoder decodes the input $x_1$ and generates outputs $b^1_i \in \{0,1\}$ for $i=0,1,\cdots,N-1$. The second decoder decodes the input $x_0$ and generates outputs $b^0_j \in \{0,1\}$ for $j=0,1,\cdots,N-1$. We will have $b^1_{x_1}$ and $b^0_{x_0}$ to be at logic level 1 for the Nary input $x_1, x_0$. All the remaining outputs will be logic level 0. Therefore, outputs of the two-to-$N^2$ decoder can be formed by logically ANDing output of the first and second decoders $b^1_{i}$ and $b^0_{j}$, respectively. Thus, only one output of the two-to-$N^2$ decoder will be at logic level 1. All other outputs will be at logic level 0 for a given input. 

We can generalize our methodology to the $M$-to-$N^M$ decoder by representing its input as $x= \sum _{j=0}^{M-1}N^jx_j$. Outputs of the decoder $b_k \in \{0,1\}$ for $k \in \{0,1,\cdots,N^M-1\}$ can be obtained by ANDing corresponding outputs of $M$ one-to-$N$ decoders. Therefore, $k$th output $b_k$ of the $M$-to-$N^M$ decoder will be

\begin{equation}\label{eqn:M_input_decoder}
  b_k= \prod_{m=0}^{M-1} b^m_i
\end{equation}

\noindent where $k \in \{0,1,\cdots,N^M-1\}$ and $b^m_i$ is the $i$th output of the $m$th 1-to-$N$ decoder. The $i$ value is calculated as $i=\mod(\lfloor k/N^{m} \rfloor, N)$ where the floor function $\lfloor \cdot \rfloor$ calculates the greatest integer less than or equal to its argument. The product operation is used to represent AND operations in Eqn.~\ref{eqn:M_input_decoder}. Output of the decoder is one-hot such that only one $b_k$ is set to logic level 1. The remaining outputs are set to logic level 0. 

We provide truth table of the two-to-nine decoder in Table~\ref{table:ternary_decoder} as an example. As can be seen in this table, we formed this decoder by using two one-to-three decoders and binary AND operations represented by the $\cdot$ sign.

\begin{table}[htbp]
	\begin{center}
	\begin{tabular}{cc|ccc|ccc|c}
		$x_1$	& $x_0$  & $b^1_2$  & $b^1_1$ & $b^1_0$ & $b^0_2$ & $b^0_1$ & $b^0_0$ & $b_k$\\
		\hline
		    0     & 0           & 0         & 0         & 1         & 0         & 0         & 1  & $b_0=b^1_0\cdot b^0_0$ \\
		    0     & 1           & 0         & 0         & 1         & 0         & 1         & 0  & $b_1=b^1_0 \cdot b^0_1$ \\
		    0     & 2           & 0         & 0         & 1         & 1         & 0         & 0  & $b_2=b^1_0 \cdot b^0_2$ \\
		    1     & 0           & 0         & 1         & 0         & 0         & 0         & 1  & $b_3=b^1_1 \cdot b^0_0$ \\
		    1     & 1           & 0         & 1         & 0         & 0         & 1         & 0  & $b_4=b^1_1 \cdot b^0_1$ \\
		    1     & 2           & 0         & 1         & 0         & 1         & 0         & 0  & $b_5=b^1_1 \cdot b^0_2$ \\
		    2     & 0           & 1         & 0         & 0         & 0         & 0         & 1  & $b_6=b^1_2 \cdot b^0_0$ \\
		    1     & 1           & 1         & 0         & 0         & 0         & 1         & 0  & $b_7=b^1_2 \cdot b^0_1$ \\
		    2     & 2           & 1         & 0         & 0         & 1         & 0         & 0  & $b_8=b^1_2 \cdot b^0_2$ \\
		\hline
	\end{tabular}
	\end{center}
	\caption{Truth table for the two-to-nine decoder.}\label{table:ternary_decoder}
\end{table}

\section{Nary Multiplexer Formation via Nary Decoder}\label{section:Nary_MUX}

We will benefit from Nary multiplexers as our second option to implement multiple-valued logic functions. Therefore, we will start with defining the Nary multiplexer in this section. The multiplexer works such that one of its Nary inputs is connected to output via its Nary select pins. Therefore, we will first propose a $N$-to-one multiplexer. While doing so, we will benefit from switches and one-to-$N$ decoder introduced in Section~\ref{sec:one_to_n_decoder}. Afterward, we will construct the $N^M$-to-one multiplexer using $N$-to-one multiplexers.

\subsection{$N$-to-One Multiplexer Formation}

\begin{figure}[htbp]
\begin{center}
\scalebox{0.75}{\includegraphics{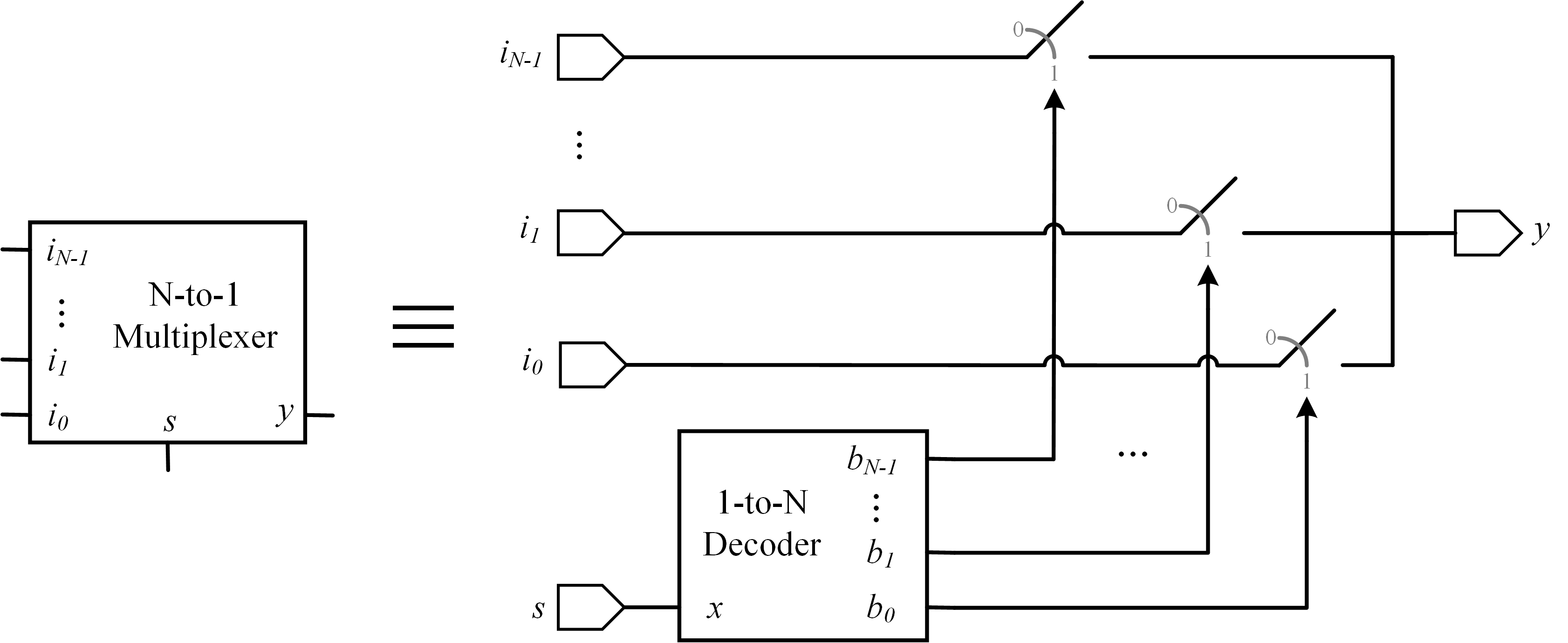}}
\end{center}
\caption{Circuit representation of the N-to-one multiplexer with one select pin. }\label{fig:Nary_Mux_circuit}
\end{figure}

Symbolic representation of the Nary $N$-to-one multiplexer with one select pin is as in Fig.~\ref{fig:Nary_Mux_circuit}. In this figure, $i_n \in \{0,1, \cdots , N\}$ for $n=0,1,\cdots,N-1$ are N separate inputs, each being an Nary number. $s$ is the Nary select input. $y$ is Nary output of the multiplexer. We can form the input output relation of the $N$-to-one multiplexer as 

\begin{equation}
  y=\sum_{n=0}^{N-1}i_n \cdot b_n
\end{equation}

\noindent where $b_n \in \{0,1\}$ are the decoded form of the select signal $s$ as explained in Section~\ref{sec:one_to_n_decoder}. To note here, only one $b_n$ will have logic level 1. Hence, we can construct the proposed multiplexer using switches and decoder as in Fig.~\ref{fig:Nary_Mux_circuit}. 

In Fig.~\ref{fig:Nary_Mux_circuit}, the one-to-$N$ decoder introduced in Section~\ref{sec:one_to_n_decoder} decides on the specific switch to be opened based on the select value $s$. Thus, the corresponding Nary input $i_n$ will be connected to output $y$ of the multiplexer. 

\subsection{$N^M$-to-One Multiplexer Formation}

\begin{figure}[htbp]
\begin{center}
\scalebox{0.75}{\includegraphics{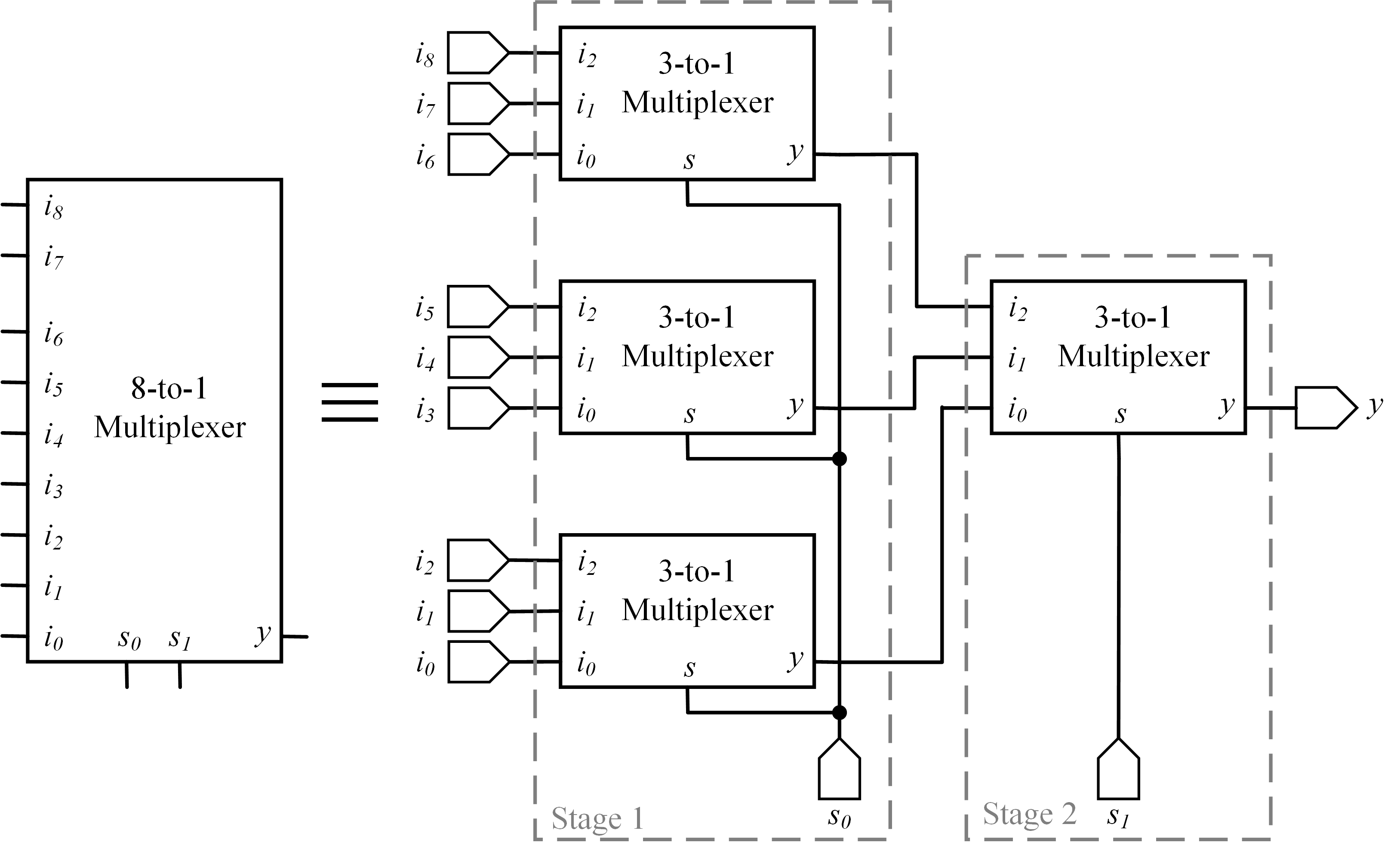}}
\end{center}
\caption{Decomposing a two select pin MUX.}\label{fig:Nary_Mux_decompose}
\end{figure}

We can form the $N^M$-to-one multiplexer by using $N$-to-one multiplexers, each having one select pin. We can explain this procedure on a simple ternary multiplexer with two select pins as in Fig.~\ref{fig:Nary_Mux_decompose}. 
 
In Fig.~\ref{fig:Nary_Mux_decompose}, the first stage of the multiplexer has three three-to-one multiplexers, each controlled by the select pin $s_0$. Output of the three multiplexers are fed to the second stage three-to-one multiplexer with the select pin $s_1$. Thus, we need four three-to-one decoders in implementation. In general, we will need $\sum_{m=0}^{M-1}N^m$ $N$-to-one multiplexers to construct the $N^M$-to-one multiplexer.

The $N^M$-to-one multiplexer can also be constructed by $N^M$ switches and one $M$-to-$N^M$ decoder. This is the direct extension of the method in Fig.~\ref{fig:Nary_Mux_circuit}. The reader can pick the suitable method for his or her own application.

\section{Multiple-Valued Logic Function Realizations}\label{section:Nary_Func}

We can realize any multiple-valued logic function by Nary decoders or multiplexers. We will consider each option separately in this section.

\subsection{Nary Decoder based Realization}

We can realize a multiple-valued logic function by Nary decoders introduced in Section~\ref{section:Nary_decoder}. To do so, we should first form truth table of the function. Then, we should group inputs yielding the same output. Afterward, we should generate a control signal by applying binary OR operation to the corresponding decoder outputs. These control signals are then used to connect the appropriate Nary value to output.

Let's consider the multiple-valued logic function $y=F(x)$ to explain the procedure proposed in this section. We can decode the Nary input $x$ as $b_n \in \{0,1\}$ for $n \in \{0,1,\cdots,N-1\}$ by our Nary decoder. We should apply binary OR to $b_n$s producing the same output depending on the truth table of $F(x)$. Output of each OR gate will be the control signal $c_n \in \{0,1\}$ for $n \in \{0,1,\cdots,N-1\}$. These control signals are then used to connect the appropriate Nary value to output. We can mathematically represent this operation as 

\begin{equation}\label{eqn:Nary_decoder_realize}
  y=\sum_{n=0}^{N-1}n \cdot c_n
\end{equation}

\begin{figure}[htbp]
\begin{center}
\scalebox{0.8}{\includegraphics{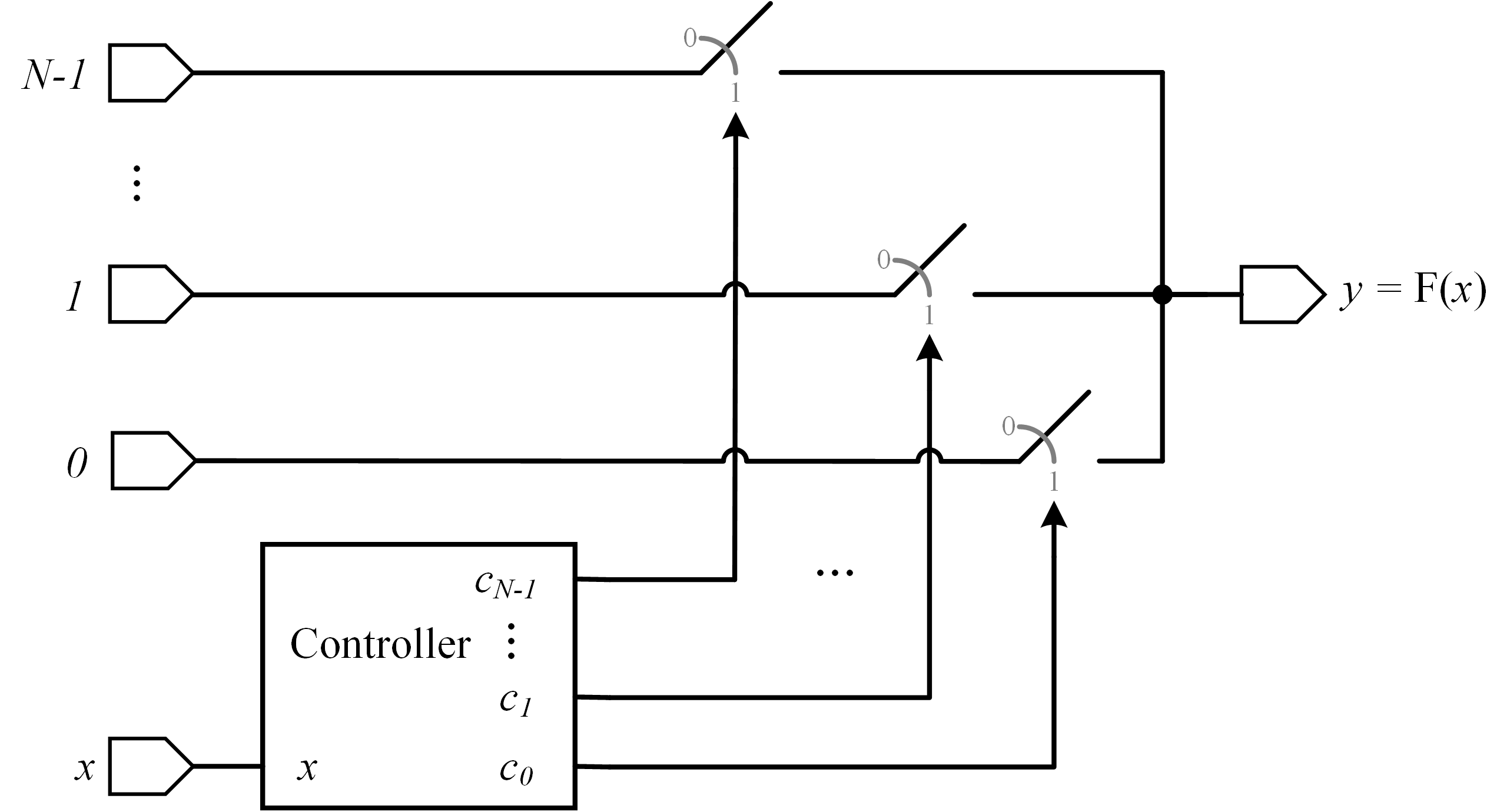}}
\end{center}
\caption{Nary decoder based realization.}\label{fig:decoder_func_realize}
\end{figure}

We can implement this structure as in Fig.~\ref{fig:decoder_func_realize}. As can be seen in this figure, there is no sum term in the implementation of Eqn.~\ref{eqn:Nary_decoder_realize}. The reason for this is that, actually only one control signal will be at logic level 1 at a time. Hence, only one of $n \in \{0,1,\cdots,(N-1)\}$ will be fed to output. 

The structure introduced in the previous paragraph can be extended to multiple-valued logic functions with $M$ inputs as $y=F(x_0, x_1, \cdots, x_{M-1})$. The only difference here will be the usage of an $M$-to-$N^{M}$ decoder in operation. The number of control signals $c_n$ will be $N$ since we will have the output $y$ in Nary logic.

We can take the ternary half-adder as an example of multiple-valued logic function. Let's realize it by the procedure introduced in this section. Therefore, we should first tabulate truth table of the half-adder as in Table~\ref{tab:truth_table_ha}. In this table, $x_1$ and $x_0$ are the two ternary numbers to be added. $b_n$ values are outputs of the two-to-nine decoder. The $sum$ and $carry$ terms represent the ternary sum and carry terms, respectively. 

\begin{table}[htbp]
	\begin{center}
	\begin{tabular}{cc|ccccccccc|cc}
		$x_1$ & $x_0$ & $b_0$ & $b_1$ & $b_2$ & $b_3$ & $b_4$ & $b_5$ & $b_6$ & $b_7$ & $b_8$ & $sum$ & $carry$\\
		\hline
		    0     & 0   & 1  & 0 & 0 & 0  & 0 & 0 & 0  & 0 & 0 & 0 & 0\\
		    0     & 1   & 0  & 1 & 0 & 0  & 0 & 0 & 0  & 0 & 0 & 1 & 0\\
		    0     & 2   & 0  & 0 & 1 & 0  & 0 & 0 & 0  & 0 & 0 & 2 & 0\\
		    1     & 0   & 0  & 0 & 0 & 1  & 0 & 0 & 0  & 0 & 0 & 1 & 0\\
		    1     & 1   & 0  & 0 & 0 & 0  & 1 & 0 & 0  & 0 & 0 & 2 & 0\\
		    1     & 2   & 0  & 0 & 0 & 0  & 0 & 1 & 0  & 0 & 0 & 0 & 1\\
		    2     & 0   & 0  & 0 & 0 & 0  & 0 & 0 & 1  & 0 & 0 & 2 & 0\\
		    1     & 1   & 0  & 0 & 0 & 0  & 0 & 0 & 0  & 1 & 0 & 0 & 1\\
		    2     & 2   & 0  & 0 & 0 & 0  & 0 & 0 & 0  & 0 & 1 & 1 & 1\\
		\hline
	\end{tabular}
	\end{center}
	\caption{Truth table of the ternary half adder.}\label{tab:truth_table_ha}
\end{table}

Based on Table~\ref{tab:truth_table_ha}, we can form $sum= 0\cdot(b_0+b_5+b_7)+1\cdot(b_1+b_3+b_8)+2\cdot(b_2+b_4+b_6)$ and $carry=1\cdot(b_5+b_7+b_8)$. Here, the binary OR operation is represented by the $+$ sign. Therefore, we can apply the binary OR operation to $b_0,b_5,b_7$ as the first group for the $sum$ term to generate $c_0$. Likewise, we can apply the binary OR operation to $b_1,b_3,b_8$ as the second group for the $sum$ term to obtain $c_1$. Finally, we can apply the binary OR operation to $b_2,b_4,b_6$ as the third group for the $sum$ term to get $c_2$. Hence, our control signals will be $c_0=b_0+b_5+b_7$, $c_1=b_1+b_3+b_8$, and $c_3=b_2+b_4+b_6$ for the ternary outputs 0, 1, and 2, respectively. Then, we can benefit from the structure as in Fig.~\ref{fig:decoder_func_realize} to realize the $sum$ term of the half adder. In a similar manner, we can form the control signal for the $carry$ term and realize it by another structure.

\subsection{Nary Multiplexer based Realization}

We can use the Nary multiplexer introduced in Section~\ref{section:Nary_MUX} to realize a multiple-valued logic function. As in the binary logic case, we should feed function values as input to the Nary multiplexer. Inputs of the function are connected to select pins of the multiplexer. Then, we will get output of the function from the multiplexer output.

As in the previous section, we can take the ternary half-adder as an example of the multiple-valued logic function. Let's consider its realization by Nary multiplexers. Since the half adder has two ternary inputs, we should use two separate nine-to-one multiplexers for the $sum$ and $carry$ terms. Let's consider realization of the $sum$ term first. Here, we will have inputs of the Nary multiplexer as $i_0=0, i_1=1, i_2=2, i_3=1, i_4=2, i_5=0, i_6=2, i_7=0, i_8=1$ based on Table~\ref{tab:truth_table_ha}. Select pins of the multiplexer will be connected to inputs of the function $x_1$ and $x_0$. Then, output of the first multiplexer can be taken as the $sum$ term. We can follow the same steps by using a second multiplexer to form the $carry$ term.   

\section{Adding Reconfigurability to the Multiple-Valued Logic Function Realizations}\label{sec:reconfigurabile_nary}

The Nary decoder introduced in Section~\ref{section:Nary_decoder} has binary output. Therefore, we can use binary D-latches in connection with decoder outputs for reconfigurable multiple-valued logic function realization. We can benefit from D-latches in a similar manner for the multiplexer based reconfigurable function realization. We will consider each option next.

\subsection{Nary Decoder based Reconfigurable Function Realization}

\begin{figure}[htbp]
\begin{center}
\scalebox{0.8}{\includegraphics{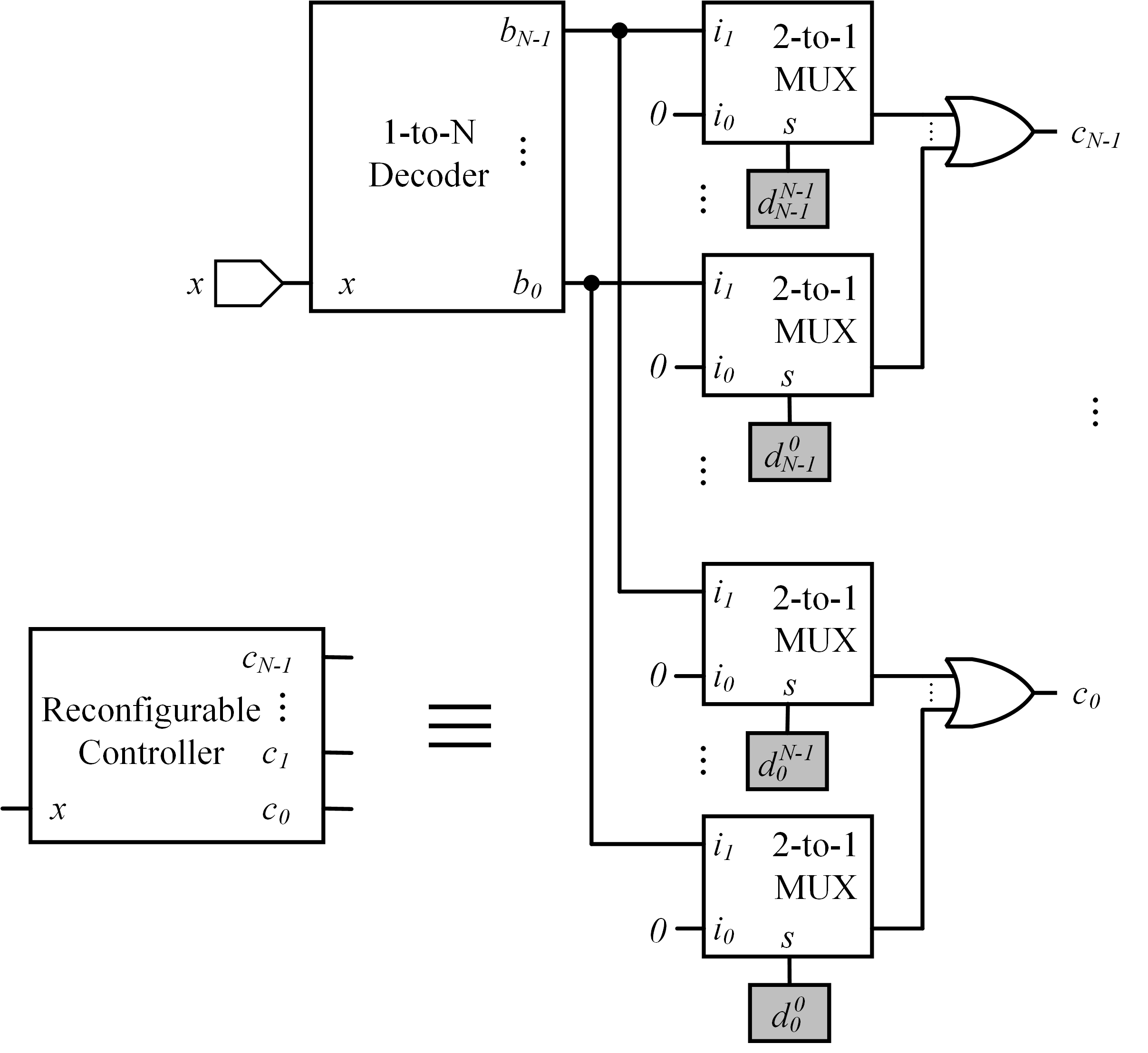}}
\end{center}
\caption{Control signal generation for the Nary decoder based reconfigurable function.}\label{fig:decoder_reconfigurable}
\end{figure}

We should add latches between the decoder and control signals to add reconfigurability to the Nary decoder based multiple-valued logic function realization. Moreover, each control signal should have all decoder inputs for reconfigurability. Hence, our control signal representation for the Nary function with one input will be $c_k=\sum_{n=0}^{N-1}b_n \cdot d^k_n$ for $k=0,\cdots, N-1$, where $d^k_n$ is the D-latch output for the $k$th control signal with input $b_n$. Here, we have the binary OR operation represented by the sum symbol. We can reconfigure the proposed structure by setting appropriate D-latch values to logic level 0 or 1. We will have a setup as in Fig.~\ref{fig:decoder_reconfigurable} in hardware form.

\subsection{Nary Multiplexer based Reconfigurable Function Realization}

\begin{figure}[htbp]
\begin{center}
\scalebox{.8}{\includegraphics{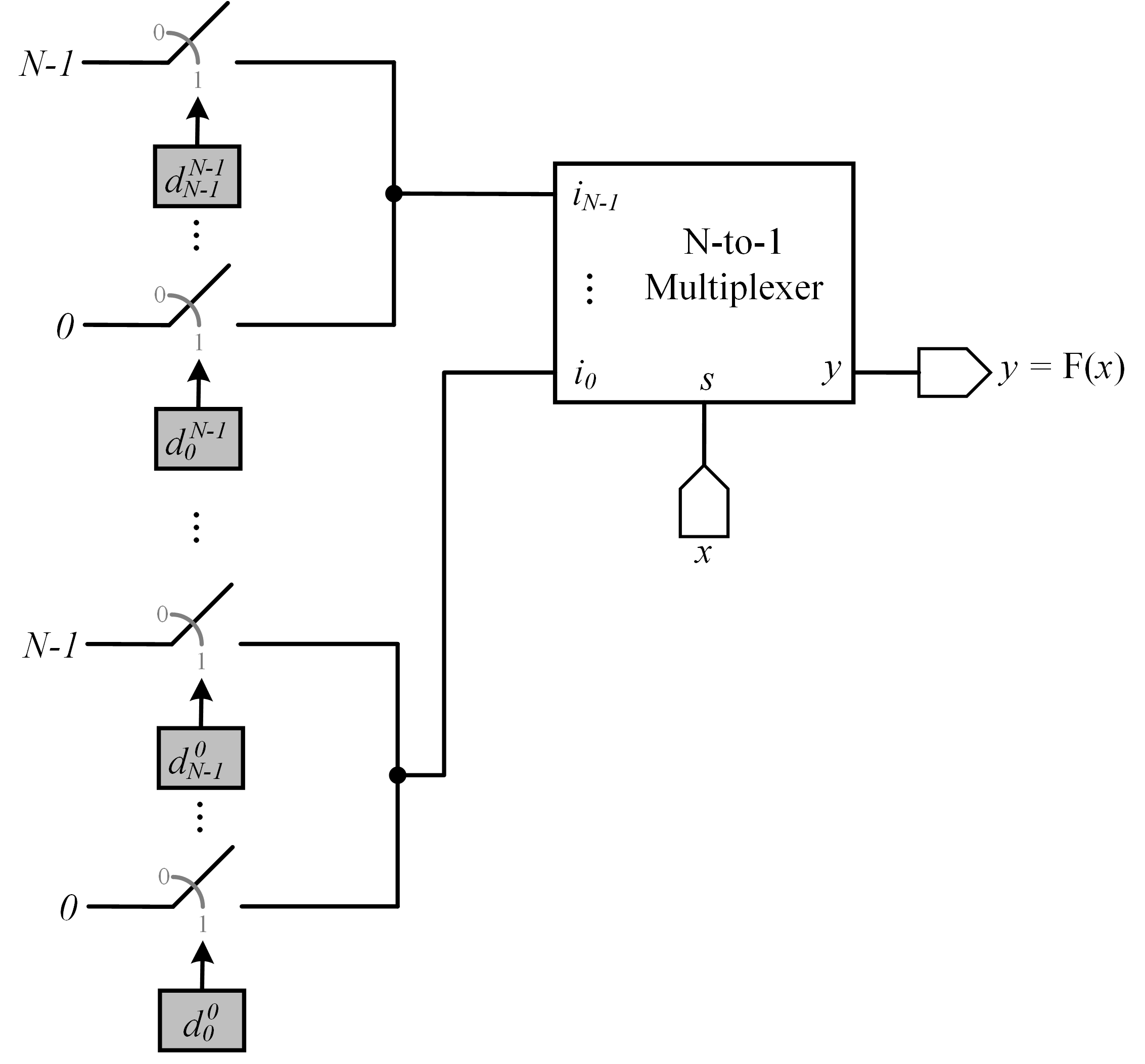}}
\end{center}
\caption{Architecture of the Nary multiplexer based reconfigurable function implementation.}\label{fig:mux_reconfigurable}
\end{figure}

We can use Nary multiplexers to realize reconfigurable Nary functions. To do so, we should add Nary selection blocks before each input of the Nary multiplexer as in Fig.~\ref{fig:mux_reconfigurable}. Therefore, $k$th input of the multiplexer will be formed as $i_k=\sum_{n=0}^{N-1}n \cdot d_n^k$ for $k=0, \cdots, N-1$, where $d_n^k$ are output of binary latches. We can reconfigure the structure by feeding logic level 0 or 1 to the corresponding D-latch input for reconfigurability.

\section{Multiple-Valued Sequential Circuit Realization}\label{sec:sequential_nary}

In this section, we cover Nary sequential circuits. First, we propose the Nary D-latch. Then, we will use it along with the Nary function realization introduced in Section~\ref{section:Nary_Func} to form Nary sequential circuits.

\subsection{Nary D-Latch}

\begin{figure}[htbp]
    \begin{center}
    	\subfigure[Nary D-latch]{\scalebox{0.8}{\includegraphics{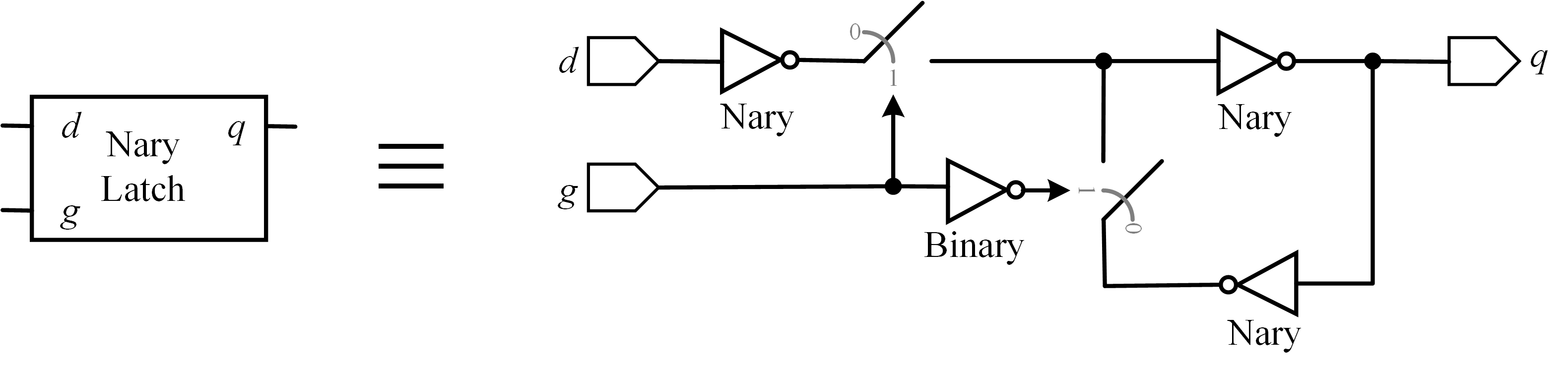}}\label{fig:Nary_latch_circuit}}
    	\subfigure[Nary D flip-flop]{\scalebox{0.8}{\includegraphics{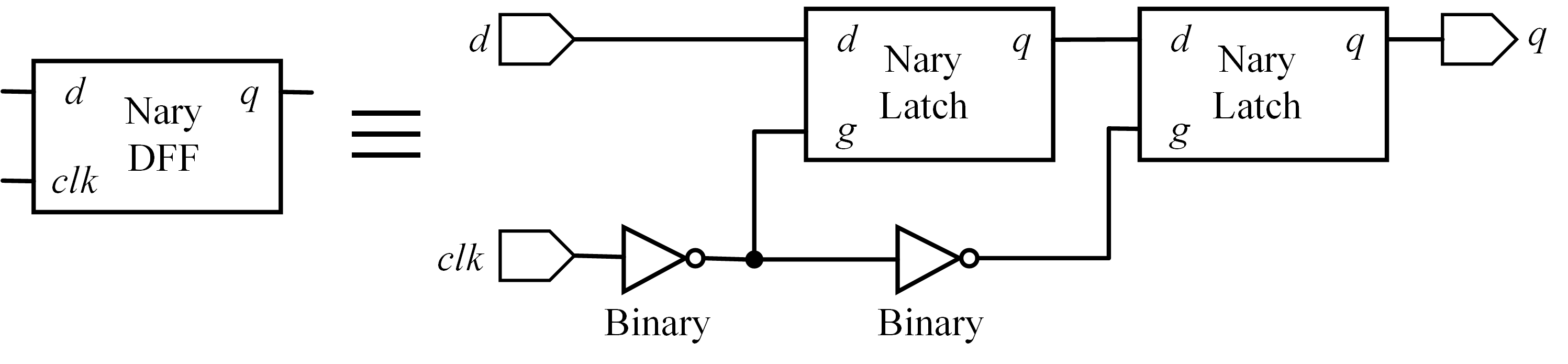}}\label{fig:Nary_flipflop_circuit}}
    \end{center}
    \caption{Circuit form of the Nary D-latch and D flip-flop.}\label{fig:Nary_latch_flipflop}
\end{figure}

We can form an Nary D-latch by realizing the formula $q=g \cdot d + g' \cdot q^{-1}$ where $d$ and $g$ stand for the data and gate pins, respectively. $q$ is the output of the Nary D-latch. $q^{-1}$ is the last value stored in the latch. When $g=0$, the output will keep its previous value. When $g \neq 0$ the output will be set to the value at the data pin $d$. We can implement the proposed Nary latch by Nary and binary inverters and switches as in Fig.~\ref{fig:Nary_latch_circuit}. If the input value to an Nary inverter is $d$, then its output will have the value $N-1-d$. The Nary inverter can be realized via one of the methods introduced in Section~\ref{section:Nary_Func}. An Nary D flip-flop can be implemented by cascading two D-latches as in Fig.~\ref{fig:Nary_flipflop_circuit}.

\subsection{Multiple-Valued Sequential Circuit Implementation}

As we have the Nary function realization procedure and D flip-flop, we can extend the binary sequential circuit design procedure for the Nary case. 

\begin{figure}[htbp]
\begin{center}
\scalebox{0.8}{\includegraphics{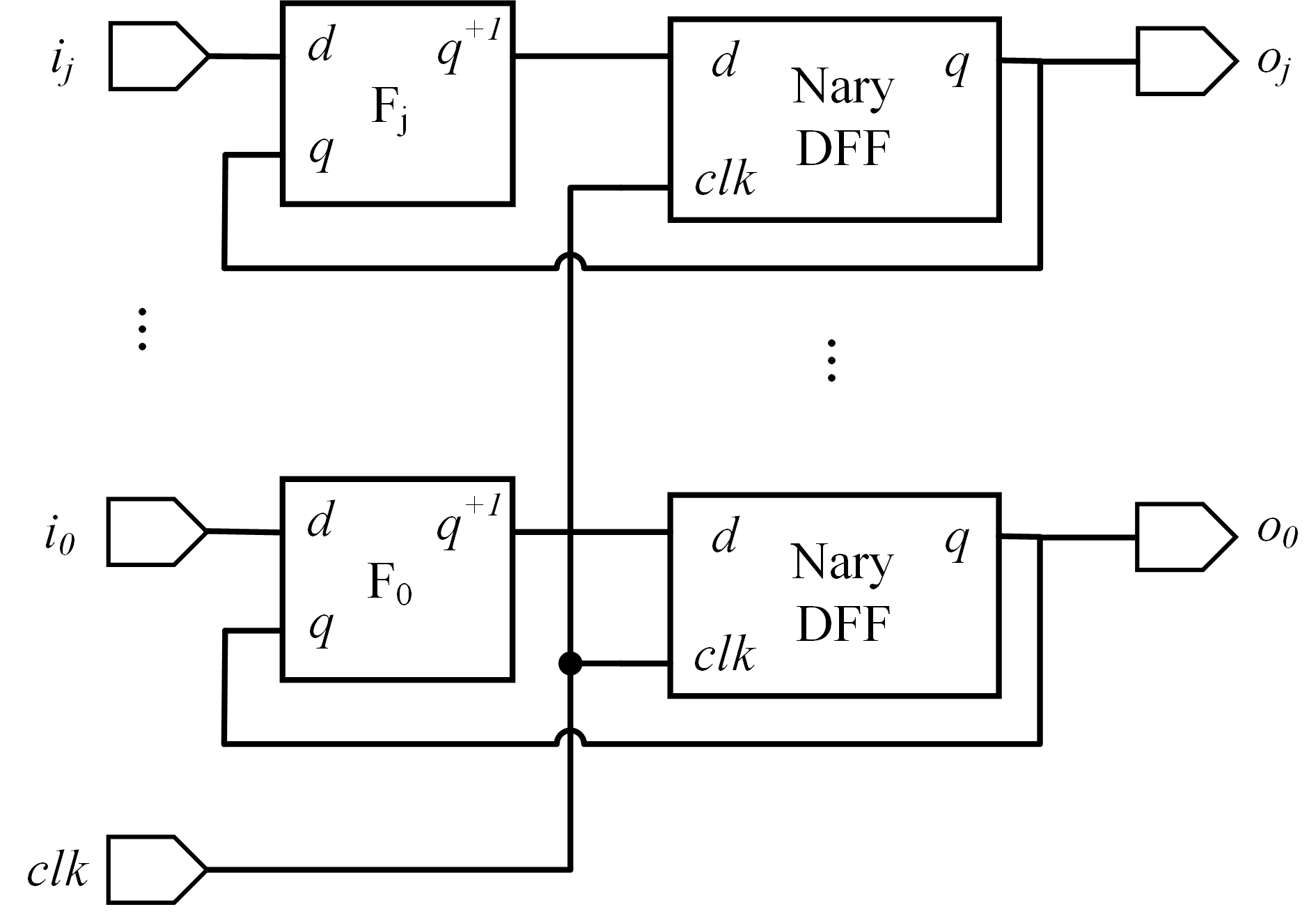}}
\end{center}
\caption{Realization of Nary sequential circuits.}\label{fig:nary_seq_cir}
\end{figure}

We can use the Nary D flip-flop and function realizations to construct sequential circuits. In general, next states $q_n^{+1}$ is calculated from the present states $q_n$ and inputs $i_n$ as $q_n^{+1}=F(q_n,i_n)$. The setup is shown in Fig.~\ref{fig:nary_seq_cir}.

\section{Final Comments} \label{sec:conclussion}

Utilization of TLG and multi-valued logic reduces power consumption and wire congestion. In this paper we introduced a general methodology to implement TLG based reconfigurable circuits. Proposed methodology may be used to build any reconfigurable multi-valued function. Modern digital designs heavily rely on design automation tools. However, similar tools are lacking for multi-valued logic designs. Development of these automation tools is a critical and open research field.

\bibliography{MVL_TLG}

\end{document}